%% file: bookchapter.tex
\documentclass[runningheads,a4paper]{llncs}

\usepackage{amssymb}
\usepackage{graphicx}
\usepackage[sort&compress,numbers]{natbib}

\usepackage{url}
\urldef{\mailsa}\path|{alfred.hofmann, ursula.barth, ingrid.haas, frank.holzwarth,|
\urldef{\mailsb}\path|anna.kramer, leonie.kunz, christine.reiss, nicole.sator,|
\urldef{\mailsc}\path|erika.siebert-cole, peter.strasser, lncs}@springer.com|    
\newcommand{\keywords}[1]{\par\addvspace\baselineskip
\noindent\keywordname\enspace\ignorespaces#1}

\begin{document}

\mainmatter  % start of an individual contribution

\title{Mental Health and Sensing}

\titlerunning{Mental Health and Sensing}

\author{Abdul Kawsar Tushar$^1$%
% \thanks{Please note that the LNCS Editorial assumes that all authors have used the western naming convention, with given names preceding surnames. This determines the structure of the names in the running heads and the author index.}%
\and Muhammad Ashad Kabir$^2$\and Syed Ishtiaque Ahmed$^1$
% \and Frank Holzwarth\and\\
% Anna Kramer\and Leonie Kunz\and Christine Rei\ss\and\\
% Nicole Sator\and Erika Siebert-Cole\and Peter Stra\ss er
}
\authorrunning{Mental Health and Sensing}

\institute{$^1$Department of Computer Science, University of Toronto, Toronto, Canada\\
$^2$School of Computing and Mathematics, Charles Sturt University, NSW, Australia\\
% \mailsa\\
% \mailsb\\
% \mailsc\\
tushar.kawsar@gmail.com, akabir@csu.edu.au, ishtiaque@cs.toronto.edu\\
% \url{http://www.springer.com/lncs}
}


\toctitle{Lecture Notes in Computer Science}
\tocauthor{Authors' Instructions}
\maketitle

\begin{abstract}
% 15-30 pages including references
Mental health is a global epidemic, affecting close to half a billion people worldwide. Chronic shortage of resources hamper detection and recovery of affected people. Effective sensing technologies can help fight the epidemic through early detection, prediction, and resulting proper treatment. Existing and novel technologies for sensing mental health state could address the aforementioned concerns by activating granular tracking of physiological, behavioral, and social signals pertaining to problems in mental health. Our paper focuses on the available methods of sensing mental health problems through direct and indirect measures. We see how active and passive sensing by technologies as well as reporting from relevant sources can contribute toward these detection methods. We also see available methods of therapeutic treatment available through digital means. We highlight a few key intervention technologies that are being developed by researchers to fight against mental illness issues.

\end{abstract}

\input{tex-files/1_introduction.tex}
\input{tex-files/2_mental_problems.tex}
\input{tex-files/3_existing.tex}
\input{tex-files/4_intervention.tex}

\input{tex-files/5_discussion.tex}

% For citations in the text please use
% square brackets and consecutive numbers: \cite{jour}, \cite{lncschap}

% \bibliographystyle{spbasic}
\bibliographystyle{spbasic_unsort}
\bibliography{bib}

\end{document}

%% file: tex-files/1_introduction.tex
\section{Introduction}

Mental health can be termed as a concern that is plaguing the entire earth. There is a worrying number of 450 million around the globe that have been diagnosed with some form of mental or neurodevelopmental illnesses \cite{brundtland2001mental} and they often lead to various levels of  disability \cite{world2003investing}. Mental and neurodevelopmental illnesses give rise to a mortality rate that has been compared at a level more than double that of the general population, leading to approximately 8 million deaths \cite{walker2015mortality}. Another impact of such sheer numbers related to these conditions is the financial burden, which generate from expenditure for care as well as the loss in productivity. The numbers related to economic loss has been estimated at more than \$400 billion dollars, that only in the United States of America for a year \cite{insel2015post}.

Well-being of patients suffering from serious mental illness for a sustainable period can be ensured through treatment, management, and care and this can be achieved through granular symptom monitoring \cite{morriss2013training}. However, existing clinical tools and resources are limited in terms of accessibility and scalability \cite{abdullah2018sensing}. Mobile health, often termed as mHealth for brevity, is where mobile (or electronic) devices converge with medical professionals and public health administration \cite{adibi2015mobile} and has been a well-researched area for exploring the scope of involving qualitative research methods with a view to providing accessible treatment, participant monitoring and retention, and progress of treatment. The growth in the number of mobile devices has also been a significant factor in lending more weight to this type of solutions, since close to 4 billion people around the world own at least one phone( the number is scheduled to double by 2022) \cite{cerwall2015ericsson}. This is remarkable when we consider the fact that studies found more than 70\% people suffering from serious mental illness have mobile phones \cite{ben2013mobile}. In addition, an increase of sensors embedded in mobile phones is giving birth to novel possibilities of utilizing these devices into mental health care based on digital evidence, such as quantitative functional and behavioral labels efficiently and without obstacles \cite{gaggioli2013mobile, mohr2017personal}. 

% A mobile phone-based approach may be valuable in gathering long-term objective data, aside from self-ratings, to predict changes in clinical states and to investigate causal inferences about state changes in patients (e.g., those with affective disorders) \cite{dogan2017smartphone}. These technologies, including active and passive measurement tools, serve to decrease geospatial, temporal, and financial barriers \cite{chan2019review}. 

% Additionally, self-monitoring has not always been shown to be a valid measurement of behavior. For example, a systematic review pointed out that electronic self-monitoring of mood among depression sufferers appeared to be a valid measure of mood in contrast to self-monitoring of mood among mania sufferers \cite{faurholt2016electronic}.

What is holding an widespread adoption of sensors in mental health care and management is the scattered and restricted state of evidence that proves the connection between, on one hand, sensor data obtained using wearables and ubiquitous smartphones and, on the other hand, the prevalence and status of mental illnesses \cite{abdullah2018sensing, or2017high}. In this paper, we show how technology can help in detection and sensing of mental health problems around the around. Specifically, we focus on the major mental illnesses that are tormenting billions of people across various countries. We see how active and passive sensing by technologies as well as reporting from relevant sources can contribute toward these detection methods. We also see available methods of therapeutic treatment available through digital means. We highlight a few key intervention technologies that are being developed by researchers to fight against mental illness issues. 

%% file: tex-files/2_mental_problems.tex
\section{Mental Health Problems}

% A brief overview of discussed mental health problems in the current world, including neurodevelopmental disorders
% \begin{enumerate}
%     \item Definition
%     \item Severity
%     \item Prevalence
% \end{enumerate}

% \textbf{Psychiatric disorders:} bipolar disorder, depression, depression symptoms, stress, symptoms of stress, schizophrenia.

In this section, we do not aim to classify mental health disorders as that is not our target for this paper. Instead, our discussion would revolve around the prevalence of these disorder to provide a sense of their different presentations. Characteristics of major mental disorders include a permutation of irregular and atypical belief, concepts, attitude, and expression especially with people in the surrounding. Top mental disorders are schizophrenia, bipolar conditions (BP), depression and sadness, and developmental disorders together with autism \cite{whomentaldisorders}. Table \ref{tab:mendis} shows the number of people affected by these disorders in the world.

% \textbf{Severity:} 

\textbf{Prevalence:} We use the World Health Organization fact sheet \cite{whomentaldisorders} to know the prevalence of these illnesses. Depression is a common psychiatric disorder that is a major concern responsible for various forms of disability. Around the world, a population of more than 200 million are affected by depression, majority of whom are women. 
% Depression can be long-lasting or recurrent, substantially impairing people’s ability to function at work or school and to cope with daily life. At its most severe, depression can lead to suicide. 
Schizophrenia can make it difficult for people affected to work or study normally. More tan 40 million humans are affected by BP, which comprises of both manic and depressive periods mingled with regular behavior and mood states. Schizophrenia is another crippling psychiatric disorder that disables around 15 million. Schizophrenia and other psychoses can distort the ability to think, perceive, and express. Another concerning illness that affects more than 50 million worldwide is dementia, which is characterized by a deterioration in cognitive function that exacerbates the regular effects of aging on memory. Lastly, developmental disorders comprise various terms such as intellectual disability and pervasive developmental disorders including autism. 1 in 160 children in the world are estimated to be affected by autism \cite{whomentaldisorders}. 

\begin{table}[!t]
\begin{center}
\renewcommand{\arraystretch}{1.5}
\begin{tabular}{c|c}
\hline
\textbf{\begin{tabular}[c]{@{}c@{}}Mental Health \\ Problem\end{tabular}} & \textbf{\begin{tabular}[c]{@{}c@{}}Severity in the\\ World\end{tabular}} \\ \hline
Depression                                                                & 264 million                                                               \\
Bipolar Disorder                                                          & 45 million                                                                \\
Schizophrenia                                                             & 20 million                                                                \\
Dementia                                                                  & 50 million                                                                \\
Autism Spectrum Disorder                                                  & 50 million                                                               
\end{tabular}
\caption{Major mental health disorders and victims of each of them in the world.}
\label{tab:mendis}
\end{center}
\end{table}

%% file: tex-files/3_existing.tex
\section{Existing Detection Techniques}

In this section, we discuss some of the widely popular methods that are used to detect the prevalence and existence of mental health problems in a person. We first discuss how the wide use of mobile phone technology is helping the revolution of mental health sensing methods. Then we discuss the currently available sensing technologies by dividing these technologies in three categories: biological or physiological sensing, behavioral or soft sensing, and social or proximity sensing methods.

% \begin{enumerate}
%     \item Biological Sensing: blood pressure, emotion change in face (Tanzeem Choudhury, Mashfiqui Rabbi) - sensors in smartphones, sensors in wearables, 
%     \item Soft Sensing: pattern recognition from usage of phone or application (Munmum De Choudhury, Saeed Abdullah) - phone lock/unlock and charge pattern, 
%     \item Social Sensing: mental health of a person is judged by people in their social circle
% \end{enumerate}

\subsection{Usage of Mobile Phones}

The wide availability of mobile and sensor technology has enabled the opportunity for personalized data collection efficiently and without obstacles.

% Decreased physical activity, abnormal sleep patterns, and social interactions are well-documented symptoms of depression \cite{american2014desk}. Significant changes in the amount of physical and mental activity can be a behavioral indicator of manic and depressive states in bipolar disorder \cite{goodwin2007manic}. Currently these conditions are measured using subjective questionnaires that have well-identified limitations \cite{stone2003patient}.

\textbf{Real-world behavior sensing:} A popular technique of estimating the status of one's mental health is to use smartphones sensors to capture behavioral data of humans. We can use a plethora of sensors available in today's smartphones. The aforementioned sensors can be used in different permutations to capture a wide range of human behavior, including mental and physical. A predecessor to this approach was used along with various sensors to capture and classify data of physical activity \cite{lester2006practical}. Another similar study was utilized to predict social isolation in older adults using sensors and body microphone \cite{rabbi2011passive}. 

\textbf{Technology-mediated behavior:} More and more work is exploring and analyzing online behavior to predict and care for mental health. For instance, a recent study by de Choudhury et al. demonstrated using information from social media to predict the early signs of depression \cite{de2013predicting}.

\subsection{Physiological Signals}
\begin{enumerate}
    \item \textbf{Facial expression:} Facial expressions and associated emotions can convey a person’s emotional states and could be useful in diagnosing psychiatric illnesses \cite{gur2006flat}. Tron et al. tracked facial Action Units using 3D cameras in people with schizophrenia to differentiate patients from control participants \cite{tron2015automated}. Another study used a similar range of technology to detect the onset of suicidal thoughts in a person \cite{laksana2017investigating}. There are other similar studies \cite{valstar2014avec, wang2015using}.
    % A related research has proposed using facial features along with audio cues for detecting levels of symptom severity in patients with depression \cite{valstar2014avec}. 
    These ideas can be paired with the widespread use of mobile phones to diagnose mental illness. 
    
    % For instance, Wang et al. developed a system that opportunistically captures one’s facial expressions throughout the day by using the front camera of a phone \cite{wang2015using}.
    
    \item \textbf{Heart rate variability:} People affected with psychiatric conditions suffer from a high chance of facing cardiovascular morbidity compared to the general population \cite{hennekens2005schizophrenia, weiner2011cardiovascular, pratt1996depression}. A recent research argued that the reduction in heart rate variability (HRV) could be behind the connection between this heightened cardiac mortality and psychiatric illness \cite{kemp2010impact}. If that is indeed true, then there is a link between HRV and depression and this could be detected through sensing. The same relationship is reported for people with post-traumatic stress disorder \cite{tan2011heart}, anxiety disorders \cite{chalmers2014anxiety}, and bipolar disorder and schizophrenia \cite{quintana2016reduced}. Traditional HRV measurement devices are heavy; therefore, smart devices can be an apt replacement. Some devices are already in place for this purpose \cite{bai2018comparative}.
    
    \item \textbf{Eye movement:} Data captured from fine changes in the movement of eyes can be used to draw conclusion and inferences about mental illness, as shown in research on schizophrenia \cite{levy2010eye} and depression \cite{winograd2006ocular, alghowinem2013eye}. A technology called EOG glasses (Electrooculography in Wearable form as in glasses) can be used for detecting the movement of eyes and well as the frequency and pattern of blinking \cite{dhuliawala2016smooth}. Additionally, web cameras can also be used for similar purposes to detect the beginning of dementia \cite{cano2017towards}.
    
    \item \textbf{Electrodermal activity:} Electrodermal activity (or EDA for short) can be gauged from measuring changes in electrical properties in human skin \cite{critchley2002electrodermal}. Various research has utilized this technique in mental health detection setting to utilize the proposed relationship between heightened EDA and mental illness \cite{schell2005electrodermal}. For instance, regarding BP, EDA signals can classify various mood states and subsequent swings \cite{lanata2014pattern, greco2014electrodermal}. EDA can also be used to determine suicidal tendencies \cite{jandl2010suicide, thorell2013electrodermal}.
    
\end{enumerate}

\subsection{Behavioral/Soft Signals}
\begin{enumerate}
    \item \textbf{Mobility and location:} Patterns in our location can be used to predict social activities, which can provide a peek into a person's mental status. A research tried to link lethargic lifestyle with depression \cite{roshanaei2009longitudinal}, while another linked phone location data to depression severity \cite{canzian2015trajectories}. There are various such other researches (see \cite{saeb2015mobile, wang2016crosscheck}).
    % A recent research used location information of a pe data to calculate mobility features \cite{saeb2015mobile}, including circadian movement, entropy, and variance, Wang et al. computed a rich set of location and mobility features using GPS data (such as total distance traveled, maximum displacement from the home, location entropy, and location routine index) that were strongly correlated with disease symptoms in patients with schizophrenia \cite{wang2016crosscheck}, and Chow et al. used similar features from GPS data for monitoring social anxiety symptoms \cite{chow2017using}.
    
    \item \textbf{Speech patterns:} Characteristics of speech could be used to judge mental health. There is an established relation between depression and human voice \cite{nilsonne1987acoustic}, as well as speech monotone \cite{moore2004comparing}. A recent research found that jitters in voice is important to establish a pattern that can identify patients having suicidal thoughts \cite{ozdas2004investigation}. Three other research has done significant work in developing frameworks to collect audio data for similar purposes \cite{faurholt2016electronic, lu2012stresssense, muaremi2014assessing}. 
    % Lu et al. developed \textit{StressSense} to collect audio data for monitoring stress in daily life \cite{lu2012stresssense}. Muaremi et al. used phone call conversation data to identify manic and depressive states in bipolar disorder \cite{muaremi2014assessing}. Faurholt-Jepsen et al. also collected voice features from phone calls and found that these features can be used to determine phases in bipolar disorder \cite{faurholt2016voice}. Rabbi et al. developed a smartphone framework for ``on the fly'' continuous collection and processing of audio data to infer presence of human voice and conversation using metrics such as spectral regularity and energy \cite{rabbi2011passive}.
    
    \item \textbf{Technology use:} The patterns demonstrated in our day-to-day technology usage contain meaningful data and this can be leveraged in the fight against mental illnesses. For instance, patterns of phone use have been linked with people's behaviors related to sleeping and waking up, something that has been explored in relation to screen lock information of modern smartphone devices \cite{abdullah2014towards}. These features were discovered to be significantly correlated with mood in BP \cite{frost2013supporting}.
    
    Other features pertaining to the pattern of usage of smartphone by a person have been found helpful. Two very recent papers explored the connection between change of technology usage and different mood states of BP patients \cite{matthews2017double, alvarez2014tell}. Schizophrenia patients also found to display similar connection \cite{wang2016crosscheck}, as did patients of depression \cite{saeb2015mobile}.
    
    \item \textbf{Activity:} While the connection between physical activity and mental state may not seem obvious, research has actually established a strong link between them. For instance, in BP, mania and depression are strongly related with overactivity and under-activity, respectively \cite{wolff1985motor}. This information can be utilized to detect and predict the onset, existence, and duration of either of these mood states in a person. Studies earlier used \textit{actigraph} to determine level of activity in a person \cite{walther2015physical, john2012actigraph}. However, different smartphone sensors could be utilized in order to monitor an user activity level on a continuous basis \cite{wu2012classification}. Studies in recent years have looked at data of numerous physical activities to connect them with mental disorders line BP \cite{osmani2013monitoring, beiwinkel2016using}, depression \cite{canzian2015trajectories}, and schizophrenia \cite{wang2016crosscheck}.
\end{enumerate}

\subsection{Social/Proximity Signals}
\begin{enumerate}
    \item \textbf{Social interaction:} Sensing technologies that can capture data from nearby access points as well as other sensing devices are helpful in obtaining information about social interactions without much obstructions. Although not as accurate as user-filled data, these social proximity data can be used as a replacement which can serve to indicate the level of social interactions of a person with the help of a good algorithm \cite{aharony2011social, ben2015mobile}. 
    
    \item \textbf{Communication patterns:} A person's level of social engagement can be indicated by their communication technology usage data. Communication patterns of humans have been used to identify various mental illnesses such as BP \cite{gur2006flat} and schizophrenia \cite{faurholt2016behavioral, wang2016crosscheck}. They found that symptoms are associated with a change of frequency and duration in outgoing SMS or phone calls.
    
    \item \textbf{Social media:} Since many individuals are comfortable in sharing their daily life stories on different social media channels, these channels can be a rich source of data to determine social engagement, social network characteristics, mood, and emotion that are related to one’s overall well-being. Recent research has worked with Twitter data to predict depression and PTSD \cite{de2013predicting, reece2017forecasting}. Images and videos in social media posts can be used to collect useful signals about a person. For example, mood \cite{abdullah2015collective} and depression \cite{reece2017instagram} can be predicted from sentiment visible in social media photos. This idea has the potential to be scalable in this age of vast social media user engagement.
    
\end{enumerate}

%% file: tex-files/4_intervention.tex
\section{Design Interventions}

In this section, we explore some of the available intervention techniques that are associated with digital sensing of mental health problems. There are various methods that are used to assist people in judging and caring for their mental health conditions. These technologies can range from using machine learning methods to detect mental illness to providing customized and personalized feedback or treatment routine. In the end, we discuss two popular methods that seamlessly integrate the sensing, inference, and partial treatment procedures. 

% \begin{enumerate}
%     \item For cure from a mental condition
%     \item For keeping the mental condition under control
%     \item Inference using ML \& DL
%     \item (1) Aggregate data (2) broad suggestions
%     \item Persuasion based on behavioral theory
%     \item Nudge theory
%     \item Correlation between diverse aspects (Bluetooth proximity with University means runny nose?) - some data gathered through questionnaire
% \end{enumerate}

If we want to a scientific model of how humans behave from the data that we capture from various wearable and mobile sensing devices, we need to develop one or more algorithms that will serve as a bridge between what we know and what we want to predict. This is useful for medical doctors as well as public health administrators alike to have useful insights about the spread and depth of mental illnesses. While statistical modeling can be used as for this purpose, they may not capture the high level of complexity present in human behavioral data \cite{breiman2000randomizing}. For this reason, researchers have been opting for machine learning methods \cite{lane2011bewell}. For instance, depression \cite{canzian2015trajectories, saeb2015mobile} and social rhythm metric \cite{abdullah2016automatic} has been inferred using smartphone data that was passively acquired. Specifically, deep learning has shown success in determining or predicting about the onset of mental illness \cite{martinez2014deep}.

\subsection{Summary and Suggestion Display}

Usually, there are two approaches for presenting the captured data using wearable and mobile sensors to the user for useful insights and feedback in the mHealth domain. First, we can summarize the collected data into short bursts of information or visual statistics or both (e.g., UbiFit \cite{consolvo2008activity} and BeWell \cite{lane2011bewell}). Second, based on the collected data, we can predict a set of present and future states and based on this prediction, we can provide recommendations to the user. While the first method usually allows for the user to set their goal based on the presented statistics, the second method would often set a goal for the user and provide recommendation to reach that goal with minimum effort and maximum efficiency. The issue is that it is difficult to make maximum or even desired use of the complexity present in the capture data. Hence, the feedback or recommendation presented to people are only partially useful and even sometimes partially correct. Even more, because of the lack of insight into the data, the recommendations can even be dangerous sometimes. The good news is that researchers have been exploring different frameworks that would formalize the process of providing feedback and recommendation that would streamline the process and minimize risks. Additionally, researchers have explore variation in feedback in the forms of data change, augmentation, or subtraction to perceive the resulting change in user behavior \cite{costa2016emotioncheck}. Below, we discuss two such frameworks.

\subsection{Frameworks for Extracting Useful Information into Recommendations and Feedback}

\begin{figure}[!t]
    \centering
    \includegraphics[width=\linewidth]{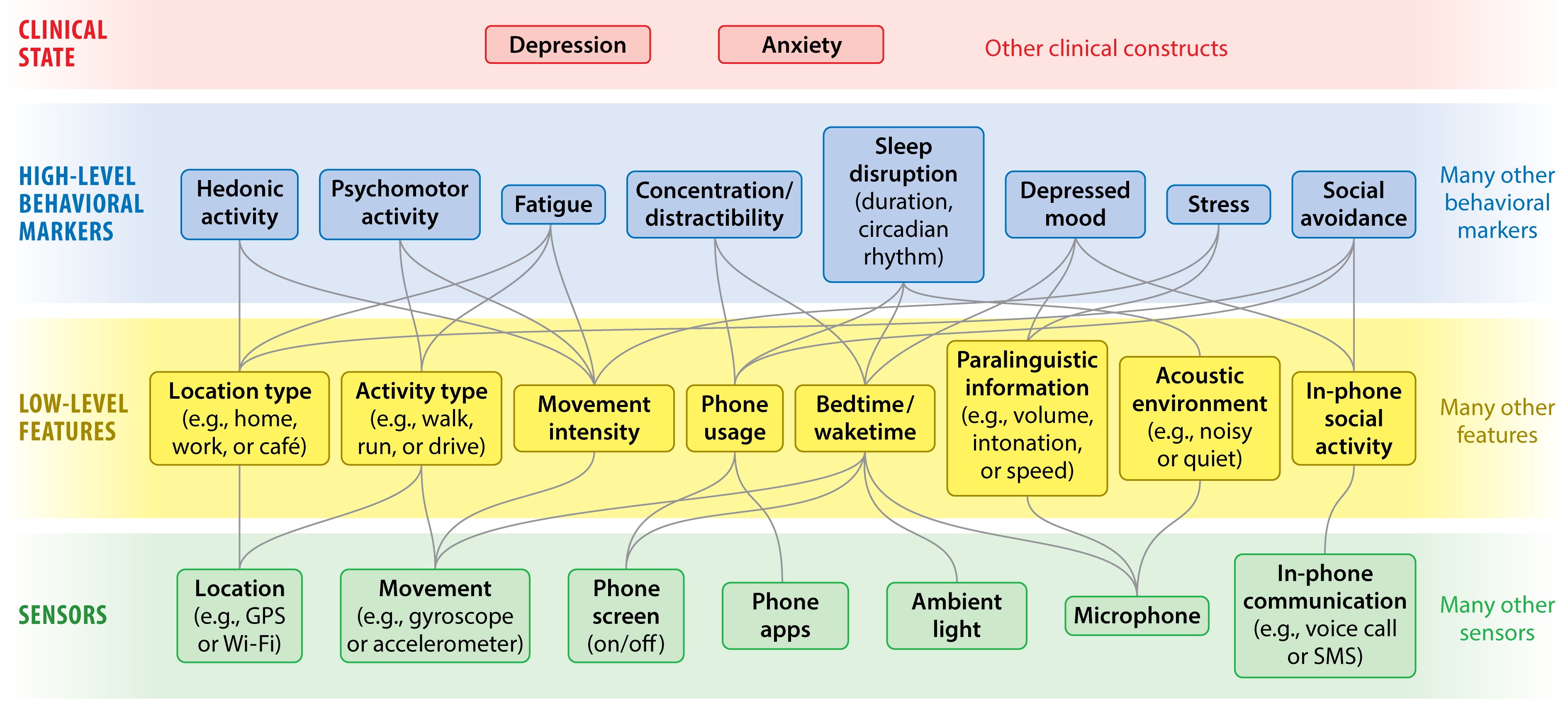}
    \caption{Example of a layered, hierarchical sensemaking framework used in \cite{mohr2017personal}. Boxes in blue are high-level behavioral markers. Boxes in yellow are features. Boxes in green are inputs to the sensing platform.}
    \label{fig:sensingstructure}
\end{figure}

A hierarchical framework extracts data from sensors and extract useful features in two sequential steps. This framework is depicted in Figure \ref{fig:sensingstructure}. The first level at the bottom (in green) has the raw inputs to the sensor device which is often a mobile phone. This data needs to be processed to extract any useful information. The second level (in yellow) is where the system merges statistical algorithms--expert in finding patterns in data-- with human intelligence via brainstorming and relevant domain expertise to construct low-level features. The boxes at the top (in blue) combine the middle-level features into behavioral markers by using machine learning and deep learning.

\begin{figure}[!t]
    \centering
    \includegraphics[width=\linewidth]{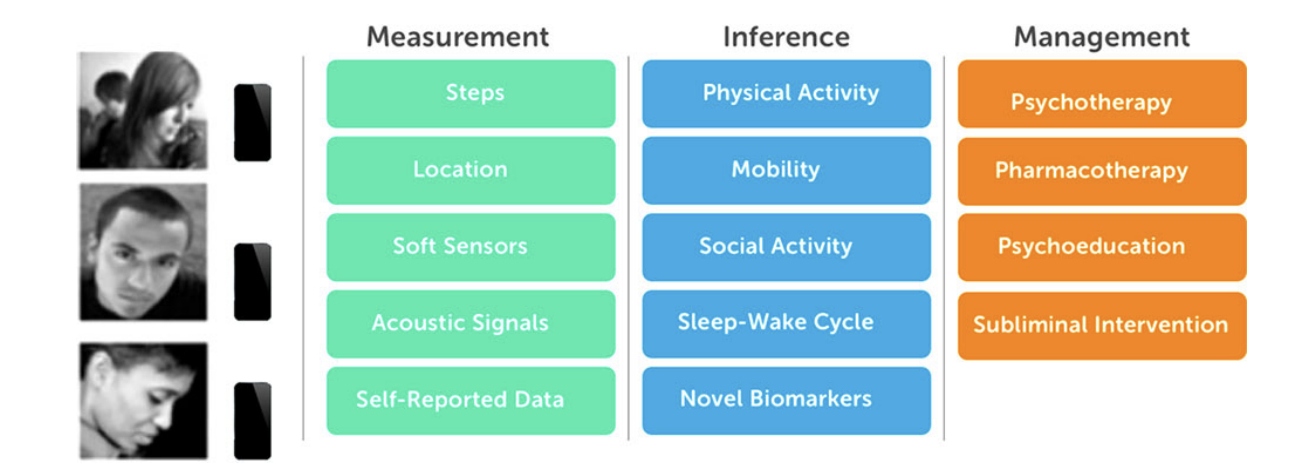}
    \caption{Framework for using behavioral sensing for detection of psychiatric illness health used in \cite{aung2017sensing}}
    \label{fig:framework}
\end{figure}

Aung et al. \cite{aung2017sensing} propose a more comprehensive and inclusive three-stage framework for integrating behavior sensing into the domain of mental health. This model can take into account user data from various physical sensors as well as self-reported data. Figure \ref{fig:framework} shows the entire process. The first stage takes into account data from sensors as well as self-reported procedures over a time to become more inclusive. Then, this raw data is processed using machine learning or other methods to create new features and gain useful insights into the data. The final phase works on integrating the inferences from the middle step toward the management of the condition if detected. This could include using the inferences alongside traditional mental health therapies to create personalized digital interventions and chrono-therapeutic interventions. 

%% file: tex-files/5_discussion.tex
\section{Discussion}

% \begin{enumerate}
%     \item Summary so far
%     \item Problems of self reporting and other current methods: issues with ML models, issues with how data is interpreted, which Behavioral marker are included, differences among humans and how the benchmark is changed/fixed (subjective measurement), lack of contextual and societal (economic/cultural) differences, privacy breach (advertisement), 
%     \item Limitations of the current classification (unclear boundary between social and soft sensing) 
%     \item Future directions
% \end{enumerate}

The existing technologies for sensing mental health condition suffer from a number of limitations. First, studies do not demonstrate a significant level of effectiveness for and correlation with improving patient detection and care for mental health patients. The dearth of clinical evidence can be linked back to the insufficient amount of research, lack of significant sample size or population, an insignificant time over which studies were live (both in short term and medium term), as well as a lack of funding opportunities, which can again lead to a lack of awareness regarding the severity and spread of these problems in different societies. These issues can be addressed by improving the study design and procedures and seeking more funding \cite{abdullah2018sensing}. Increasing awareness about mental health problems is another wing of solving these issues.

Second, not many studies strive to combine the effects of various types of sensing. If data is being collected about a single patient or the same group of patients and the result inferred from these sources are consistent, then the system could be more confident about its inference. The challenge of combining data from different streams is chiefly computational, where machine learning systems could provide to be a helping hand. However, machine learning methods also suffer from the lack of reproducibility \cite{asselbergs2016mobile}. These techniques used for mental health sensing suffer from a lack of expiration date as well as the curse of variability \cite{mohr2017personal}. Additionally, how the errors associated with the predictions of a machine learning system will be explained, addressed, and incorporated in the future iterations of the same model is something that needs more significant research, which is presently absent. Such user-facing errors need to be addressed properly and with a convincing methodology, lest we should face affecting the quality of the experience of people using technology for sensing..

Third, the nature of sensing technologies dictate that they will need to obtain a gigantic amount of data that is related to personal use and has a potential to sensitive in nature (especially when the data is related to personal behavior or health problems). Furthermore, wearable sensing technologies are not advanced enough to differentiate between target participant and non-participants and hence, due to the availability and ease of use in any setting, can risk obtaining unexpected, irrelevant, and unnecessary data \cite{ahmed2017digital} that can lead to false diagnosis, false triggers, and lack of care for patients. Researchers should strive to identify potentially sensitive data items to deploy a plan of action of protecting this data from malicious parties. In addition, researchers should also think about how they can identify unexpected, irrelevant, and unnecessary data and separate valuable data from all the clutter.

% Fourth, while sensing technologies often collect data in a passive manner, user engagement can be crucial for ensuring data quality over a long period of time. For example, technologies that use smartphone sensors need to make sure that a user carries her phone and charges it regularly. Long-term user adherence is also critical for collecting ground truth data. However, most recent studies on sensor technologies do not focus on user engagement and adherence. As a result, large-scale deployment of these technologies in the real world might face some serious challenges \cite{mohr2017personal}.

Fourth, related to the previous concern is another that is the issue of privacy and security for passively collected data. Unfortunately, the existing research community members are in a disagreement regarding what to do about this risk \cite{shilton2016we}. A crucial aspect that may often get overlooked is the difficulty in stripping identifying information from data collected by sensors in mobile phone and wearables. Even seeming innocuous data such as location information \cite{de2013unique}, when obtained in sufficient quality, can expose the identification of a victim, exposing to a risk of facing social stigma \cite{sweeney2000simple}. While there are numerous existing techniques to help de-identify data (especially location data), they are not full-proof in preserving privacy as well as usefulness of data at the same time \cite{brush2010exploring}.

% A growing cloud of digital exhaust is emitted from our daily activities and actions. Some of these data are produced intentionally, such as through the use of wearables. But much of the data are a by-product of our daily actions captured through our smartphones, computers, purchasing, and the increasingly sensor-enabled objects in our lives. The promise for research into mental health, as well as for clinical care, is enormous. But the challenges are also large and manifold. These challenges must be overcome before it is viable for digital solutions for mental health to be clinically deployed.

While an exponentially increasing amount of data is being captured from mobile and wearable sensors around the world, there are other, less structured challenges that can pose significant barriers toward capturing and treating mental illness using mobile devices. In Global South, mental health is a vastly neglected domain rife with misconceptions, maltreatment, and lack of treatment options \cite{tushar2020stigma}. Sometimes, people consult the local witches instead of medical doctors to treat illnesses that seem unusual to the community (like mental illness) \cite{sultana2019witchcraft}. Rural areas especially suffer from the lack of medical infrastructure and mental health facilities \cite{sultana2019parar} and if we plan to offset this restriction by employing mobile devices, we face another different set of challenges \cite{ahmed2017digital, ahmed2017privacy}. Furthermore, the low rate of literacy is often a hurdle for people to use smartphones \cite{ahmed2013ecologies}. These challenges should also be considered along with the more ``high-tech'' obstacles described above if we want to capture a holistic picture of the domain where we want to treat and manage psychiatric illnesses with sensors.